\documentclass[groupedaddress,
  aip,
  pop,
  amsmath,amssymb,
  reprint,
]{revtex4-1}
\usepackage{graphicx,color}
\usepackage{amsmath}
\usepackage{natbib}
\usepackage{epsfig}
\begin{document}

\title{Sound velocities of Lennard-Jones systems near the liquid-solid phase transition}

\author{Sergey A. Khrapak}
\affiliation{Institut f\"ur Materialphysik im Weltraum, Deutsches Zentrum f\"ur Luft- und Raumfahrt (DLR), 82234 We{\ss}ling, Germany;\\Bauman Moscow State Technical University, 105005 Moscow, Russia;\\ Joint Institute for High Temperatures, Russian Academy of Sciences, 125412 Moscow, Russia}


\date{\today}

\begin{abstract}
Longitudinal and transverse sound velocities of Lennard-Jones systems are calculated at the liquid-solid coexistence using the additivity principle. The results are shown to agree well with the ``exact'' values obtained from their relations to excess energy and pressure. Some consequences, in particular, in the context of the Lindemann's melting rule and Stokes-Einstein relation between the self-diffusion and viscosity coefficients are discussed. Comparison with available experimental data on the sound velocities of solid argon at melting conditions is provided.    
\end{abstract}

\maketitle

\section{Introduction}

The high frequency (instantaneous) elastic moduli and the corresponding sound velocities are important characteristics of condensed fluid and solid  phases, which affect and regulate wave propagation,~\cite{Trachenko2015} the instantaneous Poisson's ratio,~\cite{KhrapakPRE2019} the coefficient in the Stokes-Einstein relation,~\cite{ZwanzigJCP1983,KhrapakMolPhys2019} the Lindemann melting rule,~\cite{BuchenauPRE2014,KhrapakPRR2019} as well as some other simple melting rules,~\cite{BilgramPR1987} the relaxation time in the shoving model,~\cite{DyrePRE2004,DyreJCP2012} just to give few examples.

Several useful empirical observations exist. For example, it is well known that the ratio of the sound to thermal velocity of many liquid metals and metalloids has about the same value $\simeq 10$ at the melting temperature.~\cite{IidaBook,BlairsPCL_2007,RosenfeldJPCM_1999} A close value ($\simeq 9.5$) has been reported from experiments with solid argon at the melting temperature.~\cite{IshizakiJCP1975} Values close to $\simeq 9$ have also been reported  for solid hydrogen and deuterium along the melting curve.~\cite{LiebenbergPRB1978} Rosenfeld pointed out that this ``quasi-universal'' property is also shared by the hard-sphere (HS) model.~\cite{RosenfeldJPCM_1999} More recently, it has been demonstrated that this property is also exhibited by the purely repulsive soft inverse-power-law (IPL) model in a wide range of IPL exponents.~\cite{KhrapakJCP2016} An important related question is how the presence of long-range attraction (in addition to short-range repulsion) can change this picture.

It has been recently demonstrated that the Lindemann's criterion of melting can be re-formulated for two-dimensional (2D) classical solids using statistical mechanics arguments.~\cite{KhrapakPRR2019} With this formulation the expressions for the melting temperature are equivalent in three dimensions (3D) and 2D. 
An important consequence of this formulation is that the ratio of the transverse sound velocity to the thermal velocity is predicted to have a quasi-universal value along the melting curve (which can be different in 3D and 2D). This condition has been verified on soft repulsive interactions (in particular, using the Yukawa or exponentially screened Coulomb potential) in both 2D and 3D, where it works reasonably well.~\cite{KhrapakPRR2019,KhrapakJCP2018} A natural question arises whether this condition remains valid and useful also for potentials with long-ranged attraction.                   

Motivated by these and related questions, we have investigated in detail how the longitudinal and transverse sound velocities behave at the liquid-solid phase transition of 3D Lennard-Jones (LJ) systems. The high-frequency (instantaneous) velocities have been calculated using standard expressions along the liquidus and solidus of LJ system. To facilitate the calculations, we elaborate on the principle of additivity of the melting and freezing curves put forward by Rosenfeld.~\cite{RosenfeldCPL1976,RosenfeldMolPhys1976}  We are then able to quantify the behavior of sound velocities at the liquid-solid phase transition and make comparison with purely repulsive soft sphere systems. Towards the end of the paper we will also present a comparison with available experimental data on sound velocities of solid argon at melting conditions.

\section{Materials and Methods}

\subsection{Formulation}

For an arbitrary spherically symmetric pairwise repulsive potential $\phi(r)$, the longitudinal and transverse velocities can be expressed (in 3D) as follows~\cite{Schofield1966,BalucaniBook,Takeno1971}
\begin{equation}\label{long}
mc_l^2=\frac{1}{30}\sum_j\left[2r_j\phi'(r_j)+3r_j^2\phi''(r_j)\right],
\end{equation}   
and
\begin{equation}\label{trans}
mc_t^2=\frac{1}{30}\sum_j\left[4r_j\phi'(r_j)+r_j^2\phi''(r_j)\right].
\end{equation}   
Here $c_{l}$ and $c_t$ are longitudinal and transverse elastic sound velocities, $m$ is the particle mass,  the sums run over all neighbours of a given particle, and primes denote derivatives of the interaction potential with respect to the distance $r$. 

The representation of sound velocities used above is based on the relations between the sound velocities and elastic moduli, $m
\rho c_l^2=M$ and $m\rho c_t^2=G$, where $\rho$ is the particle number density. In solids $M$ and $G$ are the conventional longitudinal and shear elastic moduli and the summation is over ideal lattice sites at zero temperature. For simplicity we assume that sound velocities are isotropic. This essentially corresponds to some effective sound velocities, averaged over the directions of propagation. The bulk elastic modulus is $K=M-\tfrac{4}{3}G$. Elastic modes softening at finite temperatures is not accounted for in this formulation. In fluids, $M$ and $G$ correspond to the so-called infinite frequency (instantaneous) longitudinal and shear moduli~\cite{ZwanzigJCP1965} (often denoted as $M_{\infty}$ and $G_{\infty}$) and the summation should be performed using the actual liquid structure. This summation is usually replaced by integration involving the radial distribution function,~\cite{ZwanzigJCP1965,Schofield1966,Hubbard1969} $\sum_j(...)\rightarrow 4\pi\rho \int(...)r^2g(r)dr$, where $g(r)$ is the radial distribution function. For our present purposes summation just keeps the notation somewhat more compact. In the liquid state additional kinetic terms should appear in Eqs.~(\ref{long}) and (\ref{trans}), but these are numerically small near the liquid-solid coexistence and are therefore omitted for simplicity.   

In the plasma-related context, expressions (\ref{long}) and (\ref{trans}) with integration instead of summations are familiar as the quasi-localized charge approximation.~\cite{RosenbergPRE1997,GoldenPoP2000,KalmanPRL2000,DonkoJPCM2008,
KhrapakPoP2016}

Important thermodynamic properties of a system of interacting particles are the internal energy and pressure. For pairwise interactions the excess (over the ideal gas) contributions to the energy, $u_{\rm ex}$, and pressure, $p_{\rm ex}$, can be expressed via summations similar to those used above
\begin{equation}
u_{\rm ex}=\frac{1}{2T}\sum_j\phi(r_j),
\end{equation} 
and 
\begin{equation}\label{pex}
p_{\rm ex} = -\frac{1}{6T}\sum_jr_j\phi'(r_j).
\end{equation}
Reduced units have been used,  $u_{\rm ex}=U_{\rm ex}/NT$, $p_{\rm ex}=P_{\rm ex}/\rho T$, where the temperature $T$ is measured in energy units. Trivial manipulation with Eqs.~(\ref{long}) and (\ref{trans}) allows us to obtain the relation between the sound velocities and the excess pressure
\begin{equation}\label{Cauchy}
(c_l/v_{\rm T})^2-3(c_{t}/v_{\rm T})^2=2p_{\rm ex},
\end{equation}     
where $v_{\rm T}=\sqrt{T/m}$ denotes the thermal velocity. Eq.~(\ref{Cauchy}) is known as Cauchy relation, it is valid independently of whether kinetic terms are included or not (simply because they cancel out if retained).~\cite{ZwanzigJCP1965}

For the Lennard-Jones potential we can also express the sound velocities in terms of $u_{\rm ex}$ and $p_{\rm ex}$.~\cite{ZwanzigJCP1965} The corresponding expressions are
\begin{equation}\label{rel1}
c_l^2/v_{\rm T}^2=-\frac{72}{5}u_{\rm ex}+11 p_{\rm ex},
\end{equation}
and
\begin{equation}\label{rel2}
c_t^2/v_{\rm T}^2=-\frac{24}{5}u_{\rm ex}+3 p_{\rm ex}.
\end{equation}
The Cauchy relation is obviously satisfied.

\subsection{Inverse-power-law model}

As an important reference case let us first consider the soft-sphere model near the fluid-solid phase transition.~\cite{KhrapakJCP2016}  
The IPL potential is defined as
\begin{equation}
 \phi (r)=\epsilon(\sigma/r)^{n},
\end{equation}  
where  $\epsilon$ and $\sigma$ are the energy and length scales, and $n$ is the IPL exponent. In this case the sound velocities are directly related to the excess pressure of the system. The corresponding relations are directly obtained from Eqs.~(\ref{long}), (\ref{trans}), and (\ref{pex}) and read  
\begin{equation}\label{cl}
c_{l}^2=\frac{3n+1}{5}v_{\rm T}^2p_{\rm ex}
\end{equation}
\begin{equation}\label{ct}
c_{t}^2=\frac{n-3}{5}v_{\rm T}^2p_{\rm ex}.
\end{equation}
Note that for $n<3$ the transverse sound velocity does not become negative, because in this regime the neutralizing background should be included, which makes the excess pressure negative.~\cite{DubinPRB1994} We will not consider this regime.

\begin{figure}
\includegraphics[width=8cm]{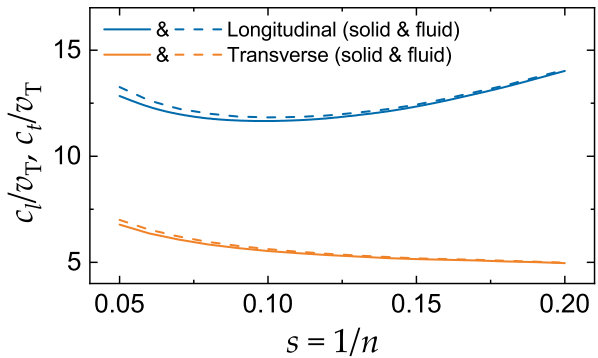}
\caption{Reduced longitudinal and transverse sound velocities at the fluid-solid coexistence of the soft sphere model versus the softness parameter $s$. Sound velocities are expressed in units of the thermal velocity $v_{\rm T}$. Upper curves are for the longitudinal mode, lower curves are for the transverse mode. Solid (dashed) curves correspond to the solid (fluid) boundary of the fluid-solid coexistence.}
\label{Fig1}
\end{figure}

The sound velocities have been evaluated along the fluid-solid coexistence boundaries using the coexistence properties tabulated in Ref.~\onlinecite{Agrawal_1995}. The inverse IPL exponent, referred to as the softness parameter $s=1/n$ varies in the range $0.05\leq s\leq 0.2$. The solid near melting is in the bcc phase for sufficiently soft interactions with $s\gtrsim 0.16$ and forms an fcc solid otherwise. Since the thermodynamic properties of the bcc crystal at melting are very nearly the same as those of the fcc crystal,~\cite{Agrawal_1995} it suffices to perform calculations for the fcc-fluid phase transition. The results are shown in Fig.~\ref{Fig1}. The following main trends are observed: (i) Very weak dependence of both longitudinal and transverse sound velocities on the softness parameter in the considered range of softness; (ii) Numerical values are comparable to those of the hard-sphere fluids at the freezing packing fraction;~\cite{KhrapakPRE2019} (iii) The difference between the sound velocities in the fluid and solid phases is very tiny and can normally be neglected; (iv) the longitudinal sound velocity exhibits a minimum at $s\simeq 0.1$, while the transverse sound velocity decreases continuously as $s$ increases; (v) The ratio of sound velocities, $c_t/c_l$, decreases monotonously from $\simeq 0.53$ to $0.35$ as $s$ increases in the range considered.   

Note that we cannot trace the transition to the HS limit using Eqs.~(\ref{cl}) and (\ref{ct}). They predict divergence of sound velocities as $n\rightarrow \infty$ (or $s\rightarrow 0$), which contradicts finite values in the HS limit.~\cite{MillerJCP1969,BrykJCP2017}  
The origin behind the unphysical divergence of the conventional expressions for the instantaneous elastic moduli when approaching the
HS limit has been identified and discussed.~\cite{KhrapakSciRep2017,KhrapakPRE2019} The conventional expressions should not be applied for $n\gtrsim 20$.

\section{Results}

\subsection{Additivity of melting curves} 

The Lennard-Jones potential is 
\begin{equation}
\phi(r)=4\epsilon\left[\left(\frac{\sigma}{r}\right)^{12}-\left(\frac{\sigma}{r}\right)^{6}\right], 
\end{equation}
where  $\epsilon$ and $\sigma$ are again the energy and length scales (or LJ units), respectively. The density, temperature, pressure, and energy expressed in LJ units are $\rho_*=\rho\sigma^3$, $T_*=T/\epsilon$, $P_*= P\sigma^3/\epsilon$, and $u_*=U/N\epsilon$. Relation to the reduced excess units is straightforward: $p_{\rm ex}=P_*/\rho_*T_*-1$ and $u_{\rm ex} = u_*/T_*-3/2$.

It has been long known, from the results of Monte Carlo (MC) simulations, that details of the interaction potential have relatively little effect on the structure of fluids near the melting temperature, in particular when extreme cases of HS and Coulomb interactions are excluded from consideration.~\cite{HansenMolPhys1973} The same concerns some simple melting characteristics such as the Lindemann ratio, reduced free volume, amplitude of the first maximum of the structure factor. These observations allowed Rosenfeld to formulate his principle of additivity of the melting curves.~\cite{RosenfeldCPL1976,RosenfeldMolPhys1976} This principle states that if the stable pairwise interaction potential represents a linear combination of stable repulsive potentials, then the temperature and pressure along the liquidus and solidus can also be expressed as a linear combinations of temperatures and pressures corresponding to individual repulsive potentials at freezing and melting.~\cite{RosenfeldCPL1976} Further support to the Rosenfeld's point of view is provided  by the concept of isomorphs,~\cite{SchroderJCP2014} which are the curves along which structure and dynamics in properly reduced units are invariant to a good approximation.~\cite{DyreJPCB2014,GnanJCP2009} Many simple systems, including the LJ case exhibit isomorphism. Melting and freezing curves appear as approximate (although not exact) isomorphs,~\cite{CostigliolaPCCP2016,PedersenNatCom2016,HeyesJCP2015} and this can simplify considerably calculation of system properties at melting and freezing.

The fact that the melting and freezing curves are approximate isomorphs indicates that structures are nearly invariants when properly scaled units are used. For instance, if the distance is measured in units of characteristic interparticle separation $a$, $\tilde{r}=r/a$, then the sums $\sum_j(1/\tilde{r}_j^n)$ are independent of density to a good approximation. This implies
\begin{equation}
\sum_j\left(\frac{\sigma}{r_j}\right)^n=\rho_*^{n/3}\Sigma_n.
\end{equation}
For an ideal zero-temperature crystal, $\Sigma_n$ would correspond to the lattice sum of the IPL-$n$ potential.
Using this property the expressions for the excess reduced energy, pressure, and sound velocities can be written in the compact form as
\begin{equation}\label{X}
{\mathcal X}={\mathcal C}_{12}\frac{\rho_*^4}{T_*}\Sigma_{12}-{\mathcal C}_6\frac{\rho_*^2}{T_*}\Sigma_{6},
\end{equation} 
where ${\mathcal X}$ is the required quantity and ${\mathcal C}_{12}$ and ${\mathcal C}_6$ are the corresponding numerical coefficients. These coefficients are summarized in Table~\ref{Tab1}. Eq.~(\ref{X}) can be referred to as additivity of excess pressure, energy, and sound velocities along melting and freezing curves.  

\begin{table}
\caption{\label{Tab1} Coefficients  ${\mathcal C}_{12}$ and ${\mathcal C}_6$ from Eq.~(\ref{X}). 
}
\begin{ruledtabular}
\begin{tabular}{lcccc}
${\mathcal X}=$ & $u_{\rm ex}$ & $p_{\rm ex}$ & $c_l^2/v_{\rm T}^2$ & $c_t^2/v_{\rm T}^2$   \\ \hline
${\mathcal C}_{12}$ & 2 & 8  & $\frac{296}{5}$ & $\frac{72}{5}$     \\
${\mathcal C}_{6}$   & 2 & 4 & $\frac{76}{5}$ & $\frac{12}{5}$      \\
\end{tabular}
\end{ruledtabular}
\end{table}   

\begin{table}
\caption{\label{Tab2} Solid-liquid coexistence data~\cite{TanMolPhys2011} and numerical values of the sums $\Sigma_n$ for $n=12$ and $n=6$ IPL potentials. 
}
\begin{ruledtabular}
\begin{tabular}{lccccc}
$n$ & $P_*$ & $\rho_{*}^{\rm sol}$ & $\rho_*^{\rm liq}$ & $\Sigma^{\rm sol}$ & $\Sigma^{\rm liq}$   \\ \hline
$n=12$ & 23.74 & 1.211  & 1.167 & 4.325 & 5.214     \\
$n=6$   &105.0 & 2.358 & 2.330 & 7.829 & 8.106      \\
\end{tabular}
\end{ruledtabular}
\end{table}  

A scaling, analogous to that of Eq.~(\ref{X}), was discussed in Refs.~\onlinecite{HeyesJCP2015,HeyesPSS2015}. From numerical data it was observed that $\Sigma_{12,6}$ vary noticeably with density close to the triple point for both phases. Away from the triple point, they become almost constant. We have chosen to take the sums $\Sigma_{12,6}$ as constant and to evaluate them from the known coexistence data for IPL $n=12$ and $n=6$ potentials.~\cite{Agrawal_1995,TanMolPhys2011} The results are summarized in Table~\ref{Tab2}. This procedure can be straightforwardly generalized to the case of a general LJ ($m-n$) potential.   

\begin{figure}
\includegraphics[width=8cm]{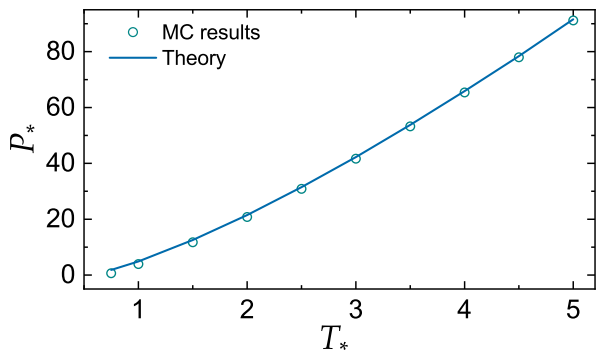}
\caption{Reduced coexistence pressure, $P_*=P\sigma^3/\epsilon$, of the Lennard-Jones system versus the reduced temperature $T_*=T/\epsilon$. Symbols correspond to MC results.~\cite{SousaJCP2012} The curve is calculated using the additivity principle (\ref{X}).}
\label{Fig2}
\end{figure}

\begin{figure}
\includegraphics[width=8cm]{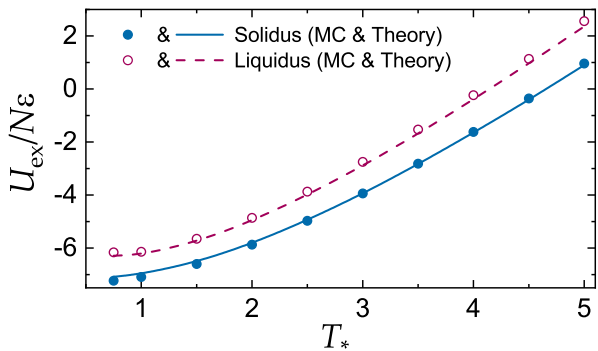}
\caption{Reduced excess energy per particle in LJ units, $U_{\rm ex}/N\epsilon$, of the Lennard-Jones system versus the reduced temperature $T_*=T/\epsilon$. Symbols correspond to the MC results.~\cite{SousaJCP2012} The curves are calculated using the additivity principle. Solid symbols and curve correspond to the solid side of the liquid-solid coexistence (solidus); open symbols and the dashed curve correspond to its liquid side (liquidus). }
\label{Fig3}
\end{figure}

Figures~\ref{Fig2} and \ref{Fig3} demonstrate the accuracy of the additivity principle (\ref{X}) when applied to calculate the reduced pressure and excess energy of the LJ systems at the fluid-solid coexistence. MC data are those tabulated in Ref.~\onlinecite{SousaJCP2012}.
Note that $P_*$ is constant at the coexistence, while $p_{\rm ex}$ is not, because density is used in the normalization. We have used solid densities and $\Sigma^{\rm sol}$ to plot the curve in Fig.~\ref{Fig2}. Very similar results would be obtained with liquid densities and $\Sigma^{\rm liq}$. Overall, the agreement between MC results and theory is rather convincing.     

\subsection{Sound velocities of the LJ system}

\begin{figure}
\includegraphics[width=8cm]{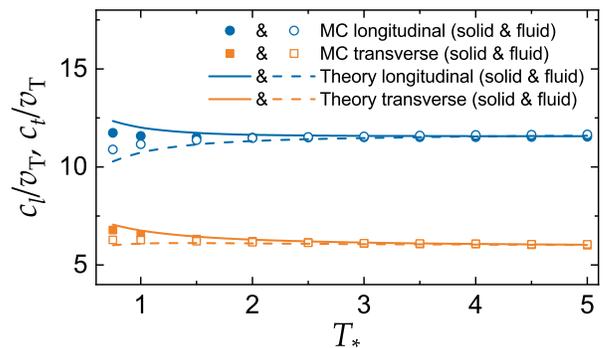}
\caption{Reduced longitudinal and transverse sound velocities of the LJ model versus the reduced temperature $T_*$. Upper symbols and curves are for the longitudinal mode, lower symbols and curves are for the transverse mode. Symbols are the results of calculation using relations (\ref{rel1}) and (\ref{rel2}) using MC data from Ref.~\onlinecite{SousaJCP2012}. Solid and open symbols correspond to the boundaries of the solid and liquid phases, respectively. Solid (dashed) curves correspond to the solid (liquid) coexistence boundary and are plotted using the  additivity principle (\ref{X}).}
\label{Fig4}
\end{figure}

The sound velocities of the LJ system at the liquid-solid coexistence are plotted in Fig.~\ref{Fig4}. Curves are calculated using the additivity principle (\ref{X}). Symbols correspond to relations (\ref{rel1}) and (\ref{rel2}) using the MC data for excess energy and pressure.~\cite{SousaJCP2012} The agreement between these two methods is good everywhere, except in the vicinity of the triple point. This is not so surprising, because in this region the two large terms associated with IPL-12 repulsion and IPL-6 attraction are almost comparable in magnitude, so that even a small relative inaccuracy in each term can result in a much greater relative inaccuracy of their difference. 

The reduced sound velocities are to a good approximation constant there with $c_l/v_{\rm T}\simeq 11.5$ and $c_t/v_{\rm T}\simeq 6$. The ratio of sound velocities is approximately constant with $c_t/c_l\simeq 0.5$.  
The difference between the sound velocities at the solid and liquid coexistence boundaries is vanishingly small away from the triple point, similarly to the IPL case. 

Now we can elaborate on some consequences of the obtained results. The presence of long range attraction seems not to affect (or, more precisely, affects rather weakly) the sound velocities in the vicinity of the fluid-solid phase transition. For comparison, the sound velocities of IPL-12 system at melting is $c_l/v_{\rm T}\simeq 11.7$ and $c_t/v_{\rm T}\simeq 5.8$. For the IPL-6 system we have  $c_l\simeq 12.9$ and $c_t\simeq 5.1$.

The Lindemann's criterion of melting can be expressed as ~\cite{BuchenauPRE2014,KhrapakPRR2019}  
\begin{equation}\label{Lindemann1}
T_{\rm m}\simeq Cm\omega_{\rm D}^2a^2,
\end{equation}
where $\omega_{\rm D}$ is the Debye frequency, $a$ is the characteristic interparticle separation (the Wigner-Seitz radius is used here), and $C$ is expected to be a quasi-universal constant. In this form the Lindemann expression for the melting temperature applies to both 3D and 2D solids (the constants $C$ can be different for the 2D and 3D cases).~\cite{KhrapakPRR2019} The Debye frequency in 3D can be expressed via the longitudinal and transverse sound velocity as
\begin{equation}\label{wD1}
\omega_{\rm D}^3=18\pi^2 \rho \left(c_l^{-3}+2c_t^{-3}\right)^{-1}.
\end{equation}       
For soft repulsive interactions the strong inequality $c_l^2\gg c_t^2$ usually holds and this can be used to further simplify the melting condition.~\cite{KhrapakPRR2019} In the considered case $c_t/c_l\simeq 0.5$, and we use the actual values  $c_l/v_{\rm T}\simeq 11.5$ and $c_t/v_{\rm T}\simeq 6$ to get $\omega_{\rm D}^2\simeq 260 v_{\rm T}^2/a^2$. The resulting value $C\simeq 0.004$ is somewhat below $C\simeq 0.006$, previously reported for a one-component Coulomb plasma.~\cite{KhrapakPRR2019} On the other hand, 
constancy of either $c_l/v_{\rm T}$ or $c_t/v_{\rm T}$, combined with the scaling (\ref{X}) immediately yield freezing and melting equations of the form
\begin{equation}
T_*^{\rm L,S}={\mathcal A}^{\rm L,S}\rho_*^4+{\mathcal B}_*^{\rm L,S}\rho_*^2,
\end{equation}
which is a very robust result reproduced in a number of various theories and approximations.   
\cite{PedersenNatCom2016,RosenfeldCPL1976,RosenfeldMolPhys1976,
KhrapakJCP2011_2,HeyesJCP2015,HeyesPSS2015,KhrapakAIPAdv2016,CostigliolaPCCP2016}

Zwanzig's result for the relation between the self-diffusion and viscosity coefficients of liquids~\cite{ZwanzigJCP1983} can be expressed in the form of the Stokes-Einstein (SE) relation with a coefficient that depends on the ratio of the transverse to longitudinal sound velocities,~\cite{KhrapakMolPhys2019} 
\begin{equation}\label{DZ1}
D\eta(\Delta/T)= 0.132\left(1+\frac{c_t^2}{2c_l^2}\right)=\alpha,
\end{equation}
where $D$ is the coefficient of self-diffusion, $\Delta=\rho^{-1/3}$, $\eta$ is the viscosity, and $\alpha$ is the SE coefficient. For $c_t/c_l\simeq 0.5$ the SE coefficient becomes $0.149$, which is in remarkable agreement with recent extensive MD simulation results demonstrating that not too far from the fluid-solid coexistence $\alpha\simeq 0.146$.~\cite{CostigliolaJCP2019}

\subsection{Comparison with experiment}

\begin{figure}
\includegraphics[width=8cm]{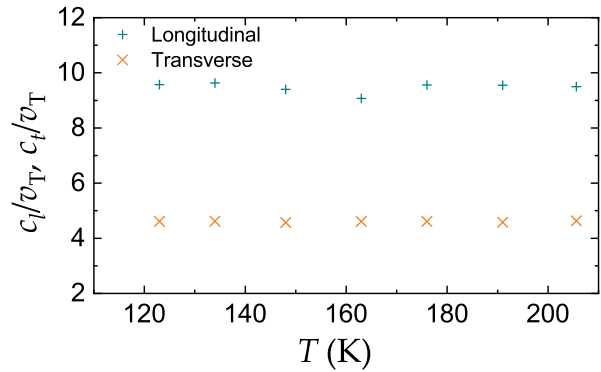}
\caption{Reduced longitudinal and transverse sound velocities in compressed solidified argon at the melting temperature in the range 
$123 - 206$ K. Symbols correspond to the experimental results tabulated in Ref.~\onlinecite{IshizakiJCP1975}.}
\label{Fig5}
\end{figure}

Measurements of the longitudinal and transverse ultrasonic wave velocities in compressed,
solidified argon for pressures up to 6 kbar (600 MPa) corresponding to melting temperatures in the range
$123-206$ K have been reported in Ref.~\onlinecite{IshizakiJCP1975}.  Experimental results, in the form of the temperature dependence of the reduced sound velocities $c_l/v_{\rm T}$ and $c_t/v_{\rm T}$, are plotted in Fig.~\ref{Fig5}. We observe that the reduced sound velocities are constant to a very good accuracy in the temperature range investigated. The ratio $c_t/c_l$ is close to 0.5 in agreement with theoretical expectations. The numerical values $c_l/v_{\rm T}\simeq 9.5$ and $c_t/v_{\rm T}\simeq 4.5$ are however by about $20\%$ lower than the theory predicts (we have used $\epsilon=125.7$ K and $\sigma = 3.345\times 10^{-8}$ cm as LJ parameters for argon~\cite{WhiteJCP1999}). In this context, we should remind that the theory of sound velocities considered here is idealized in a sense that it takes into account the temperature effect on the system structural properties, but does not take into account the effect of thermal fluctuations. The latter is known to somewhat reduce the values of elastic constants (elastic moduli softening)~\cite{Squire1969} and become important as the melting transition is approached. For many metallic solids the reduction in the shear modulus (compared to zero-temperature conditions) amounts to $20-30\%$.~\cite{Preston1992,BurakovskyPRB2003} Similarly, $\simeq 20\%$ reduction was observed in the one-component plasma model.~\cite{OgataPRA1990} Thus, the trends observed for the LJ system are not unique. Additionally, we should not ignore the fact that the pairwise Lennard-Jones potential function itself may not be the best representation of a real interaction potential in solid argon.

\section{Discussion and conclusion}

The longitudinal and transverse sound velocities of the Lenard-Jones system have been evaluated at the liquid-solid coexistence. Two methods have been employed, one uses the additivity principle and the other uses relations between sound velocities and excess energy and pressure. The first method is simple but approximate, while the second is exact, but requires the knowledge of the excess energy and pressure. The agreement between the two methods is rather good, the deviations are only observable near the triple point. This is not surprising, because in this region the repulsive and attractive contributions are comparable in magnitude, so that even a small relative inaccuracy in each of these terms can result in a much greater relative inaccuracy of their difference. Nevertheless, even near the triple point the results based on the additivity principle demonstrate acceptable accuracy.         

The calculated ratios of sound velocities to the thermal velocity are practically constant along the melting and freezing curves. The difference between the sound velocities in the solid and liquid phases is insignificant. The numerical values $c_l/v_{\rm T}\simeq 11.5$ and $c_t/v_{\rm T}\simeq 6$  are comparable with those of repulsive soft sphere  (IPL)  and HS models. Thus, long-range attraction seems not to affect sound velocities considerably. 

The latter conclusion should not be understood too literally. The structure of the LJ system in the vicinity of the fluid-solid phase transition is merely determined by the short-range repulsive branch of the interaction potential at distances about the mean inter-particle separation $\Delta$.~\cite{KhrapakJCP2011_3} In this range the potential can be approximated by the IPL shape with an effective IPL exponent $n_{\rm eff}$ (generally, $n_{\rm eff}\geq 12$).~\cite{KhrapakJCP2011_3} The value of $n_{\rm eff}$ is considerably affected by the attractive term of the LJ potential. However, since $c_l/v_{\rm T}$ and $c_t/v_{\rm T}$ are only weakly dependent on $n_{\rm eff}$ (see e.g. Fig.~\ref{Fig1}), the effect of attractive force is small in the considered context. The opposite situation is not impossible in other circumstances. For example, the presence of a long-range attractive (dipole-like) interaction can considerably suppress the sound velocity in complex plasma fluids (fluids composed of macroscopic charged particles immersed in a plasma environment).~\cite{RosenbergJPP2015,Mierk}          

The obtained results have been analysed in the context of the Lindemann's melting rule and Stokes-Einstein relation. The constancy of   $c_l/v_{\rm T}$ and  $c_t/v_{\rm T}$ is consistent with the functional form for the dependence of melting and freezing temperatures on density, which emerge in various theories and approximations. The ratio of the transverse and longitudinal sound velocities allows to evaluate the numerical coefficient in the Stokes-Einstein relation, and the result agrees remarkably well with that from recent MD simulations.

A comparison with available experimental data on the sound velocities of solid argon at melting conditions has been provided. It is demonstrated that the ratio $c_t/c_l$ in experiment is close to 0.5 in agreement with theoretical expectations. The ratios  $c_l/v_{\rm T}$ and  $c_t/v_{\rm T}$ are however by about $20\%$ lower than the theory predicts. This can be a consequence of elastic moduli softening on approaching the melting temperature, which is not taken into account in the theory. Additionally, the LJ potential may not be the best model of real interactions in solid argon.

As a final remark, we should point out that the results presented here can be easily generalized to the case of the ($m-n$) Lennard-Jones potential.
 
\acknowledgments

I would like to thank Mierk Schwabe for a careful reading of the manuscript. Work on the effect of attraction on collective modes in fluids was supported by the Russian Science Foundation, Grant No. 20-12-00356.

\bibliographystyle{aipnum4-1}
\bibliography{Additivity}

\providecommand{\noopsort}[1]{}\providecommand{\singleletter}[1]{#1}%
\begin{thebibliography}{54}%
\makeatletter
\providecommand \@ifxundefined [1]{%
 \@ifx{#1\undefined}
}%
\providecommand \@ifnum [1]{%
 \ifnum #1\expandafter \@firstoftwo
 \else \expandafter \@secondoftwo
 \fi
}%
\providecommand \@ifx [1]{%
 \ifx #1\expandafter \@firstoftwo
 \else \expandafter \@secondoftwo
 \fi
}%
\providecommand \natexlab [1]{#1}%
\providecommand \enquote  [1]{``#1''}%
\providecommand \bibnamefont  [1]{#1}%
\providecommand \bibfnamefont [1]{#1}%
\providecommand \citenamefont [1]{#1}%
\providecommand \href@noop [0]{\@secondoftwo}%
\providecommand \href [0]{\begingroup \@sanitize@url \@href}%
\providecommand \@href[1]{\@@startlink{#1}\@@href}%
\providecommand \@@href[1]{\endgroup#1\@@endlink}%
\providecommand \@sanitize@url [0]{\catcode `\\12\catcode `\$12\catcode
  `\&12\catcode `\#12\catcode `\^12\catcode `\_12\catcode `\%12\relax}%
\providecommand \@@startlink[1]{}%
\providecommand \@@endlink[0]{}%
\providecommand \url  [0]{\begingroup\@sanitize@url \@url }%
\providecommand \@url [1]{\endgroup\@href {#1}{\urlprefix }}%
\providecommand \urlprefix  [0]{URL }%
\providecommand \Eprint [0]{\href }%
\providecommand \doibase [0]{http://dx.doi.org/}%
\providecommand \selectlanguage [0]{\@gobble}%
\providecommand \bibinfo  [0]{\@secondoftwo}%
\providecommand \bibfield  [0]{\@secondoftwo}%
\providecommand \translation [1]{[#1]}%
\providecommand \BibitemOpen [0]{}%
\providecommand \bibitemStop [0]{}%
\providecommand \bibitemNoStop [0]{.\EOS\space}%
\providecommand \EOS [0]{\spacefactor3000\relax}%
\providecommand \BibitemShut  [1]{\csname bibitem#1\endcsname}%
\let\auto@bib@innerbib\@empty
\bibitem [{\citenamefont {Trachenko}\ and\ \citenamefont
  {Brazhkin}(2015)}]{Trachenko2015}%
  \BibitemOpen
  \bibfield  {author} {\bibinfo {author} {\bibfnamefont {K.}~\bibnamefont
  {Trachenko}}\ and\ \bibinfo {author} {\bibfnamefont {V.~V.}\ \bibnamefont
  {Brazhkin}},\ }\href {\doibase 10.1088/0034-4885/79/1/016502} {\bibfield
  {journal} {\bibinfo  {journal} {Rep. Progr. Phys.}\ }\textbf {\bibinfo
  {volume} {79}},\ \bibinfo {pages} {016502} (\bibinfo {year}
  {2015})}\BibitemShut {NoStop}%
\bibitem [{\citenamefont {Khrapak}(2019{\natexlab{a}})}]{KhrapakPRE2019}%
  \BibitemOpen
  \bibfield  {author} {\bibinfo {author} {\bibfnamefont {S.}~\bibnamefont
  {Khrapak}},\ }\href {\doibase 10.1103/physreve.100.032138} {\bibfield
  {journal} {\bibinfo  {journal} {Phys. Rev. E}\ }\textbf {\bibinfo {volume}
  {100}},\ \bibinfo {pages} {032138} (\bibinfo {year}
  {2019}{\natexlab{a}})}\BibitemShut {NoStop}%
\bibitem [{\citenamefont {Zwanzig}(1983)}]{ZwanzigJCP1983}%
  \BibitemOpen
  \bibfield  {author} {\bibinfo {author} {\bibfnamefont {R.}~\bibnamefont
  {Zwanzig}},\ }\href {\doibase 10.1063/1.446338} {\bibfield  {journal}
  {\bibinfo  {journal} {J. Chem. Phys.}\ }\textbf {\bibinfo {volume} {79}},\
  \bibinfo {pages} {4507} (\bibinfo {year} {1983})}\BibitemShut {NoStop}%
\bibitem [{\citenamefont {Khrapak}(2019{\natexlab{b}})}]{KhrapakMolPhys2019}%
  \BibitemOpen
  \bibfield  {author} {\bibinfo {author} {\bibfnamefont {S.}~\bibnamefont
  {Khrapak}},\ }\href {\doibase 10.1080/00268976.2019.1643045} {\bibfield
  {journal} {\bibinfo  {journal} {Mol. Phys.}\ }\textbf {\bibinfo {volume}
  {118}},\ \bibinfo {pages} {e1643045} (\bibinfo {year}
  {2019}{\natexlab{b}})}\BibitemShut {NoStop}%
\bibitem [{\citenamefont {Buchenau}, \citenamefont {Zorn},\ and\ \citenamefont
  {Ramos}(2014)}]{BuchenauPRE2014}%
  \BibitemOpen
  \bibfield  {author} {\bibinfo {author} {\bibfnamefont {U.}~\bibnamefont
  {Buchenau}}, \bibinfo {author} {\bibfnamefont {R.}~\bibnamefont {Zorn}}, \
  and\ \bibinfo {author} {\bibfnamefont {M.~A.}\ \bibnamefont {Ramos}},\ }\href
  {\doibase 10.1103/physreve.90.042312} {\bibfield  {journal} {\bibinfo
  {journal} {Phys. Rev. E}\ }\textbf {\bibinfo {volume} {90}},\ \bibinfo
  {pages} {042312} (\bibinfo {year} {2014})}\BibitemShut {NoStop}%
\bibitem [{\citenamefont {Khrapak}(2020)}]{KhrapakPRR2019}%
  \BibitemOpen
  \bibfield  {author} {\bibinfo {author} {\bibfnamefont {S.~A.}\ \bibnamefont
  {Khrapak}},\ }\href {\doibase 10.1103/physrevresearch.2.012040} {\bibfield
  {journal} {\bibinfo  {journal} {Phys. Rev. Research}\ }\textbf {\bibinfo
  {volume} {2}},\ \bibinfo {pages} {012040} (\bibinfo {year}
  {2020})}\BibitemShut {NoStop}%
\bibitem [{\citenamefont {Bilgram}(1987)}]{BilgramPR1987}%
  \BibitemOpen
  \bibfield  {author} {\bibinfo {author} {\bibfnamefont {J.}~\bibnamefont
  {Bilgram}},\ }\href {\doibase 10.1016/0370-1573(87)90047-0} {\bibfield
  {journal} {\bibinfo  {journal} {Phys. Rep.}\ }\textbf {\bibinfo {volume}
  {153}},\ \bibinfo {pages} {1} (\bibinfo {year} {1987})}\BibitemShut {NoStop}%
\bibitem [{\citenamefont {Dyre}\ and\ \citenamefont
  {Olsen}(2004)}]{DyrePRE2004}%
  \BibitemOpen
  \bibfield  {author} {\bibinfo {author} {\bibfnamefont {J.~C.}\ \bibnamefont
  {Dyre}}\ and\ \bibinfo {author} {\bibfnamefont {N.~B.}\ \bibnamefont
  {Olsen}},\ }\href {\doibase 10.1103/physreve.69.042501} {\bibfield  {journal}
  {\bibinfo  {journal} {Phys. Rev. E}\ }\textbf {\bibinfo {volume} {69}},\
  \bibinfo {pages} {042501} (\bibinfo {year} {2004})}\BibitemShut {NoStop}%
\bibitem [{\citenamefont {Dyre}\ and\ \citenamefont
  {Wang}(2012)}]{DyreJCP2012}%
  \BibitemOpen
  \bibfield  {author} {\bibinfo {author} {\bibfnamefont {J.~C.}\ \bibnamefont
  {Dyre}}\ and\ \bibinfo {author} {\bibfnamefont {W.~H.}\ \bibnamefont
  {Wang}},\ }\href {\doibase 10.1063/1.4724102} {\bibfield  {journal} {\bibinfo
   {journal} {J. Chem. Phys.}\ }\textbf {\bibinfo {volume} {136}},\ \bibinfo
  {pages} {224108} (\bibinfo {year} {2012})}\BibitemShut {NoStop}%
\bibitem [{\citenamefont {Iida}\ and\ \citenamefont
  {Guthrie}(1988)}]{IidaBook}%
  \BibitemOpen
  \bibfield  {author} {\bibinfo {author} {\bibfnamefont {T.}~\bibnamefont
  {Iida}}\ and\ \bibinfo {author} {\bibfnamefont {R.}~\bibnamefont {Guthrie}},\
  }\href@noop {} {\emph {\bibinfo {title} {The Physical Properties of Liquid
  Metals}}}\ (\bibinfo  {publisher} {Oxford University Press},\ \bibinfo {year}
  {1988})\BibitemShut {NoStop}%
\bibitem [{\citenamefont {Blairs}(2007)}]{BlairsPCL_2007}%
  \BibitemOpen
  \bibfield  {author} {\bibinfo {author} {\bibfnamefont {S.}~\bibnamefont
  {Blairs}},\ }\href {\doibase 10.1080/00319100701272084} {\bibfield  {journal}
  {\bibinfo  {journal} {Phys. Chem. Liq.}\ }\textbf {\bibinfo {volume} {45}},\
  \bibinfo {pages} {399} (\bibinfo {year} {2007})}\BibitemShut {NoStop}%
\bibitem [{\citenamefont {Rosenfeld}(1999)}]{RosenfeldJPCM_1999}%
  \BibitemOpen
  \bibfield  {author} {\bibinfo {author} {\bibfnamefont {Y.}~\bibnamefont
  {Rosenfeld}},\ }\href {\doibase 10.1088/0953-8984/11/10/002} {\bibfield
  {journal} {\bibinfo  {journal} {J. Phys.: Condens. Matter}\ }\textbf
  {\bibinfo {volume} {11}},\ \bibinfo {pages} {L71} (\bibinfo {year}
  {1999})}\BibitemShut {NoStop}%
\bibitem [{\citenamefont {Ishizaki}, \citenamefont {Spain},\ and\ \citenamefont
  {Bolsaitis}(1975)}]{IshizakiJCP1975}%
  \BibitemOpen
  \bibfield  {author} {\bibinfo {author} {\bibfnamefont {K.}~\bibnamefont
  {Ishizaki}}, \bibinfo {author} {\bibfnamefont {I.~L.}\ \bibnamefont {Spain}},
  \ and\ \bibinfo {author} {\bibfnamefont {P.}~\bibnamefont {Bolsaitis}},\
  }\href {\doibase 10.1063/1.431500} {\bibfield  {journal} {\bibinfo  {journal}
  {J. Chem. Phys.}\ }\textbf {\bibinfo {volume} {63}},\ \bibinfo {pages} {1401}
  (\bibinfo {year} {1975})}\BibitemShut {NoStop}%
\bibitem [{\citenamefont {Liebenberg}, \citenamefont {Mills},\ and\
  \citenamefont {Bronson}(1978)}]{LiebenbergPRB1978}%
  \BibitemOpen
  \bibfield  {author} {\bibinfo {author} {\bibfnamefont {D.~H.}\ \bibnamefont
  {Liebenberg}}, \bibinfo {author} {\bibfnamefont {R.~L.}\ \bibnamefont
  {Mills}}, \ and\ \bibinfo {author} {\bibfnamefont {J.~C.}\ \bibnamefont
  {Bronson}},\ }\href {\doibase 10.1103/physrevb.18.4526} {\bibfield  {journal}
  {\bibinfo  {journal} {Phys. Rev. B}\ }\textbf {\bibinfo {volume} {18}},\
  \bibinfo {pages} {4526} (\bibinfo {year} {1978})}\BibitemShut {NoStop}%
\bibitem [{\citenamefont {Khrapak}(2016)}]{KhrapakJCP2016}%
  \BibitemOpen
  \bibfield  {author} {\bibinfo {author} {\bibfnamefont {S.~A.}\ \bibnamefont
  {Khrapak}},\ }\href {\doibase 10.1063/1.4944824} {\bibfield  {journal}
  {\bibinfo  {journal} {J. Chem. Phys.}\ }\textbf {\bibinfo {volume} {144}},\
  \bibinfo {pages} {126101} (\bibinfo {year} {2016})}\BibitemShut {NoStop}%
\bibitem [{\citenamefont {Khrapak}(2018)}]{KhrapakJCP2018}%
  \BibitemOpen
  \bibfield  {author} {\bibinfo {author} {\bibfnamefont {S.}~\bibnamefont
  {Khrapak}},\ }\href {\doibase 10.1063/1.5027201} {\bibfield  {journal}
  {\bibinfo  {journal} {J. Chem. Phys.}\ }\textbf {\bibinfo {volume} {148}},\
  \bibinfo {pages} {146101} (\bibinfo {year} {2018})}\BibitemShut {NoStop}%
\bibitem [{\citenamefont {Rosenfeld}(1976{\natexlab{a}})}]{RosenfeldCPL1976}%
  \BibitemOpen
  \bibfield  {author} {\bibinfo {author} {\bibfnamefont {Y.}~\bibnamefont
  {Rosenfeld}},\ }\href {\doibase 10.1016/0009-2614(76)80048-6} {\bibfield
  {journal} {\bibinfo  {journal} {Chem. Phys. Lett.}\ }\textbf {\bibinfo
  {volume} {38}},\ \bibinfo {pages} {591} (\bibinfo {year}
  {1976}{\natexlab{a}})}\BibitemShut {NoStop}%
\bibitem [{\citenamefont
  {Rosenfeld}(1976{\natexlab{b}})}]{RosenfeldMolPhys1976}%
  \BibitemOpen
  \bibfield  {author} {\bibinfo {author} {\bibfnamefont {Y.}~\bibnamefont
  {Rosenfeld}},\ }\href {\doibase 10.1080/00268977600102381} {\bibfield
  {journal} {\bibinfo  {journal} {Mol. Phys.}\ }\textbf {\bibinfo {volume}
  {32}},\ \bibinfo {pages} {963} (\bibinfo {year}
  {1976}{\natexlab{b}})}\BibitemShut {NoStop}%
\bibitem [{\citenamefont {Schofield}(1966)}]{Schofield1966}%
  \BibitemOpen
  \bibfield  {author} {\bibinfo {author} {\bibfnamefont {P.}~\bibnamefont
  {Schofield}},\ }\href {\doibase 10.1088/0370-1328/88/1/318} {\bibfield
  {journal} {\bibinfo  {journal} {Proc. Phys. Soc.}\ }\textbf {\bibinfo
  {volume} {88}},\ \bibinfo {pages} {149} (\bibinfo {year} {1966})}\BibitemShut
  {NoStop}%
\bibitem [{\citenamefont {Balucani}\ \emph {et~al.}(1994)\citenamefont
  {Balucani}, \citenamefont {Balucani}, \citenamefont {Zoppi},\ and\
  \citenamefont {Balucani}}]{BalucaniBook}%
  \BibitemOpen
  \bibfield  {author} {\bibinfo {author} {\bibfnamefont {U.}~\bibnamefont
  {Balucani}}, \bibinfo {author} {\bibfnamefont {U.}~\bibnamefont {Balucani}},
  \bibinfo {author} {\bibfnamefont {M.}~\bibnamefont {Zoppi}}, \ and\ \bibinfo
  {author} {\bibfnamefont {Z.}~\bibnamefont {Balucani}},\ }\href@noop {} {\emph
  {\bibinfo {title} {Dynamics of the Liquid State -}}}\ (\bibinfo  {publisher}
  {Clarendon Press},\ \bibinfo {address} {Oxford},\ \bibinfo {year}
  {1994})\BibitemShut {NoStop}%
\bibitem [{\citenamefont {Takeno}\ and\ \citenamefont
  {G{\^{o}}da}(1971)}]{Takeno1971}%
  \BibitemOpen
  \bibfield  {author} {\bibinfo {author} {\bibfnamefont {S.}~\bibnamefont
  {Takeno}}\ and\ \bibinfo {author} {\bibfnamefont {M.}~\bibnamefont
  {G{\^{o}}da}},\ }\href {\doibase 10.1143/ptp.45.331} {\bibfield  {journal}
  {\bibinfo  {journal} {Progress Theor. Phys.}\ }\textbf {\bibinfo {volume}
  {45}},\ \bibinfo {pages} {331} (\bibinfo {year} {1971})}\BibitemShut
  {NoStop}%
\bibitem [{\citenamefont {Zwanzig}\ and\ \citenamefont
  {Mountain}(1965)}]{ZwanzigJCP1965}%
  \BibitemOpen
  \bibfield  {author} {\bibinfo {author} {\bibfnamefont {R.}~\bibnamefont
  {Zwanzig}}\ and\ \bibinfo {author} {\bibfnamefont {R.~D.}\ \bibnamefont
  {Mountain}},\ }\href {\doibase 10.1063/1.1696718} {\bibfield  {journal}
  {\bibinfo  {journal} {J. Chem. Phys.}\ }\textbf {\bibinfo {volume} {43}},\
  \bibinfo {pages} {4464} (\bibinfo {year} {1965})}\BibitemShut {NoStop}%
\bibitem [{\citenamefont {Hubbard}\ and\ \citenamefont
  {Beeby}(1969)}]{Hubbard1969}%
  \BibitemOpen
  \bibfield  {author} {\bibinfo {author} {\bibfnamefont {J.}~\bibnamefont
  {Hubbard}}\ and\ \bibinfo {author} {\bibfnamefont {J.}~\bibnamefont
  {Beeby}},\ }\href {\doibase 10.1088/0022-3719/2/3/318} {\bibfield  {journal}
  {\bibinfo  {journal} {J. Phys. C}\ }\textbf {\bibinfo {volume} {2}},\
  \bibinfo {pages} {556} (\bibinfo {year} {1969})}\BibitemShut {NoStop}%
\bibitem [{\citenamefont {Rosenberg}\ and\ \citenamefont
  {Kalman}(1997)}]{RosenbergPRE1997}%
  \BibitemOpen
  \bibfield  {author} {\bibinfo {author} {\bibfnamefont {M.}~\bibnamefont
  {Rosenberg}}\ and\ \bibinfo {author} {\bibfnamefont {G.}~\bibnamefont
  {Kalman}},\ }\href {\doibase 10.1103/physreve.56.7166} {\bibfield  {journal}
  {\bibinfo  {journal} {Phys. Rev. E}\ }\textbf {\bibinfo {volume} {56}},\
  \bibinfo {pages} {7166} (\bibinfo {year} {1997})}\BibitemShut {NoStop}%
\bibitem [{\citenamefont {Golden}\ and\ \citenamefont
  {Kalman}(2000)}]{GoldenPoP2000}%
  \BibitemOpen
  \bibfield  {author} {\bibinfo {author} {\bibfnamefont {K.~I.}\ \bibnamefont
  {Golden}}\ and\ \bibinfo {author} {\bibfnamefont {G.~J.}\ \bibnamefont
  {Kalman}},\ }\href@noop {} {\bibfield  {journal} {\bibinfo  {journal} {Phys.
  Plasmas}\ }\textbf {\bibinfo {volume} {7}},\ \bibinfo {pages} {14} (\bibinfo
  {year} {2000})}\BibitemShut {NoStop}%
\bibitem [{\citenamefont {Kalman}, \citenamefont {Rosenberg},\ and\
  \citenamefont {DeWitt}(2000)}]{KalmanPRL2000}%
  \BibitemOpen
  \bibfield  {author} {\bibinfo {author} {\bibfnamefont {G.}~\bibnamefont
  {Kalman}}, \bibinfo {author} {\bibfnamefont {M.}~\bibnamefont {Rosenberg}}, \
  and\ \bibinfo {author} {\bibfnamefont {H.~E.}\ \bibnamefont {DeWitt}},\
  }\href {\doibase 10.1103/physrevlett.84.6030} {\bibfield  {journal} {\bibinfo
   {journal} {Phys. Rev. Lett.}\ }\textbf {\bibinfo {volume} {84}},\ \bibinfo
  {pages} {6030} (\bibinfo {year} {2000})}\BibitemShut {NoStop}%
\bibitem [{\citenamefont {Donko}, \citenamefont {Kalman},\ and\ \citenamefont
  {Hartmann}(2008)}]{DonkoJPCM2008}%
  \BibitemOpen
  \bibfield  {author} {\bibinfo {author} {\bibfnamefont {Z.}~\bibnamefont
  {Donko}}, \bibinfo {author} {\bibfnamefont {G.~J.}\ \bibnamefont {Kalman}}, \
  and\ \bibinfo {author} {\bibfnamefont {P.}~\bibnamefont {Hartmann}},\
  }\href@noop {} {\bibfield  {journal} {\bibinfo  {journal} {J. Phys.: Condens.
  Matter}\ }\textbf {\bibinfo {volume} {20}},\ \bibinfo {pages} {413101}
  (\bibinfo {year} {2008})}\BibitemShut {NoStop}%
\bibitem [{\citenamefont {Khrapak}\ \emph {et~al.}(2016)\citenamefont
  {Khrapak}, \citenamefont {Klumov}, \citenamefont {Couedel},\ and\
  \citenamefont {Thomas}}]{KhrapakPoP2016}%
  \BibitemOpen
  \bibfield  {author} {\bibinfo {author} {\bibfnamefont {S.~A.}\ \bibnamefont
  {Khrapak}}, \bibinfo {author} {\bibfnamefont {B.}~\bibnamefont {Klumov}},
  \bibinfo {author} {\bibfnamefont {L.}~\bibnamefont {Couedel}}, \ and\
  \bibinfo {author} {\bibfnamefont {H.~M.}\ \bibnamefont {Thomas}},\
  }\href@noop {} {\bibfield  {journal} {\bibinfo  {journal} {Phys. Plasmas}\
  }\textbf {\bibinfo {volume} {23}},\ \bibinfo {pages} {023702} (\bibinfo
  {year} {2016})}\BibitemShut {NoStop}%
\bibitem [{\citenamefont {Dubin}\ and\ \citenamefont
  {Dewitt}(1994)}]{DubinPRB1994}%
  \BibitemOpen
  \bibfield  {author} {\bibinfo {author} {\bibfnamefont {D.~H.~E.}\
  \bibnamefont {Dubin}}\ and\ \bibinfo {author} {\bibfnamefont
  {H.}~\bibnamefont {Dewitt}},\ }\href {\doibase 10.1103/physrevb.49.3043}
  {\bibfield  {journal} {\bibinfo  {journal} {Phys. Rev. B}\ }\textbf {\bibinfo
  {volume} {49}},\ \bibinfo {pages} {3043} (\bibinfo {year}
  {1994})}\BibitemShut {NoStop}%
\bibitem [{\citenamefont {Agrawal}\ and\ \citenamefont
  {Kofke}(1995)}]{Agrawal_1995}%
  \BibitemOpen
  \bibfield  {author} {\bibinfo {author} {\bibfnamefont {R.}~\bibnamefont
  {Agrawal}}\ and\ \bibinfo {author} {\bibfnamefont {D.~A.}\ \bibnamefont
  {Kofke}},\ }\href {\doibase 10.1080/00268979500100911} {\bibfield  {journal}
  {\bibinfo  {journal} {Mol. Phys.}\ }\textbf {\bibinfo {volume} {85}},\
  \bibinfo {pages} {23} (\bibinfo {year} {1995})}\BibitemShut {NoStop}%
\bibitem [{\citenamefont {Miller}(1969)}]{MillerJCP1969}%
  \BibitemOpen
  \bibfield  {author} {\bibinfo {author} {\bibfnamefont {B.~N.}\ \bibnamefont
  {Miller}},\ }\href {\doibase 10.1063/1.1671437} {\bibfield  {journal}
  {\bibinfo  {journal} {J. Chem. Phys.}\ }\textbf {\bibinfo {volume} {50}},\
  \bibinfo {pages} {2733} (\bibinfo {year} {1969})}\BibitemShut {NoStop}%
\bibitem [{\citenamefont {Bryk}\ \emph {et~al.}(2017)\citenamefont {Bryk},
  \citenamefont {Huerta}, \citenamefont {Hordiichuk},\ and\ \citenamefont
  {Trokhymchuk}}]{BrykJCP2017}%
  \BibitemOpen
  \bibfield  {author} {\bibinfo {author} {\bibfnamefont {T.}~\bibnamefont
  {Bryk}}, \bibinfo {author} {\bibfnamefont {A.}~\bibnamefont {Huerta}},
  \bibinfo {author} {\bibfnamefont {V.}~\bibnamefont {Hordiichuk}}, \ and\
  \bibinfo {author} {\bibfnamefont {A.~D.}\ \bibnamefont {Trokhymchuk}},\
  }\href {\doibase 10.1063/1.4997640} {\bibfield  {journal} {\bibinfo
  {journal} {J. Chem. Phys.}\ }\textbf {\bibinfo {volume} {147}},\ \bibinfo
  {pages} {064509} (\bibinfo {year} {2017})}\BibitemShut {NoStop}%
\bibitem [{\citenamefont {Khrapak}, \citenamefont {Klumov},\ and\ \citenamefont
  {Couedel}(2017)}]{KhrapakSciRep2017}%
  \BibitemOpen
  \bibfield  {author} {\bibinfo {author} {\bibfnamefont {S.}~\bibnamefont
  {Khrapak}}, \bibinfo {author} {\bibfnamefont {B.}~\bibnamefont {Klumov}}, \
  and\ \bibinfo {author} {\bibfnamefont {L.}~\bibnamefont {Couedel}},\
  }\href@noop {} {\bibfield  {journal} {\bibinfo  {journal} {Sci. Reports}\
  }\textbf {\bibinfo {volume} {7}},\ \bibinfo {pages} {7985} (\bibinfo {year}
  {2017})}\BibitemShut {NoStop}%
\bibitem [{\citenamefont {Hansen}\ and\ \citenamefont
  {Schiff}(1973)}]{HansenMolPhys1973}%
  \BibitemOpen
  \bibfield  {author} {\bibinfo {author} {\bibfnamefont {J.-P.}\ \bibnamefont
  {Hansen}}\ and\ \bibinfo {author} {\bibfnamefont {D.}~\bibnamefont
  {Schiff}},\ }\href {\doibase 10.1080/00268977300101121} {\bibfield  {journal}
  {\bibinfo  {journal} {Mol. Phys.}\ }\textbf {\bibinfo {volume} {25}},\
  \bibinfo {pages} {1281} (\bibinfo {year} {1973})}\BibitemShut {NoStop}%
\bibitem [{\citenamefont {Schroder}\ and\ \citenamefont
  {Dyre}(2014)}]{SchroderJCP2014}%
  \BibitemOpen
  \bibfield  {author} {\bibinfo {author} {\bibfnamefont {T.~B.}\ \bibnamefont
  {Schroder}}\ and\ \bibinfo {author} {\bibfnamefont {J.~C.}\ \bibnamefont
  {Dyre}},\ }\href {\doibase 10.1063/1.4901215} {\bibfield  {journal} {\bibinfo
   {journal} {J. Chem. Phys.}\ }\textbf {\bibinfo {volume} {141}},\ \bibinfo
  {pages} {204502} (\bibinfo {year} {2014})}\BibitemShut {NoStop}%
\bibitem [{\citenamefont {Dyre}(2014)}]{DyreJPCB2014}%
  \BibitemOpen
  \bibfield  {author} {\bibinfo {author} {\bibfnamefont {J.~C.}\ \bibnamefont
  {Dyre}},\ }\href {\doibase 10.1021/jp501852b} {\bibfield  {journal} {\bibinfo
   {journal} {J. Phys. Chem. B}\ }\textbf {\bibinfo {volume} {118}},\ \bibinfo
  {pages} {10007} (\bibinfo {year} {2014})}\BibitemShut {NoStop}%
\bibitem [{\citenamefont {Gnan}\ \emph {et~al.}(2009)\citenamefont {Gnan},
  \citenamefont {Schr{\o}der}, \citenamefont {Pedersen}, \citenamefont
  {Bailey},\ and\ \citenamefont {Dyre}}]{GnanJCP2009}%
  \BibitemOpen
  \bibfield  {author} {\bibinfo {author} {\bibfnamefont {N.}~\bibnamefont
  {Gnan}}, \bibinfo {author} {\bibfnamefont {T.~B.}\ \bibnamefont
  {Schr{\o}der}}, \bibinfo {author} {\bibfnamefont {U.~R.}\ \bibnamefont
  {Pedersen}}, \bibinfo {author} {\bibfnamefont {N.~P.}\ \bibnamefont
  {Bailey}}, \ and\ \bibinfo {author} {\bibfnamefont {J.~C.}\ \bibnamefont
  {Dyre}},\ }\href {\doibase 10.1063/1.3265957} {\bibfield  {journal} {\bibinfo
   {journal} {J. Chem. Phys.}\ }\textbf {\bibinfo {volume} {131}},\ \bibinfo
  {pages} {234504} (\bibinfo {year} {2009})}\BibitemShut {NoStop}%
\bibitem [{\citenamefont {Costigliola}, \citenamefont {Schr{\o}der},\ and\
  \citenamefont {Dyre}(2016)}]{CostigliolaPCCP2016}%
  \BibitemOpen
  \bibfield  {author} {\bibinfo {author} {\bibfnamefont {L.}~\bibnamefont
  {Costigliola}}, \bibinfo {author} {\bibfnamefont {T.~B.}\ \bibnamefont
  {Schr{\o}der}}, \ and\ \bibinfo {author} {\bibfnamefont {J.~C.}\ \bibnamefont
  {Dyre}},\ }\href {\doibase 10.1039/c5cp06363a} {\bibfield  {journal}
  {\bibinfo  {journal} {Phys. Chem. Chem. Phys.}\ }\textbf {\bibinfo {volume}
  {18}},\ \bibinfo {pages} {14678} (\bibinfo {year} {2016})}\BibitemShut
  {NoStop}%
\bibitem [{\citenamefont {Pedersen}\ \emph {et~al.}(2016)\citenamefont
  {Pedersen}, \citenamefont {Costigliola}, \citenamefont {Bailey},
  \citenamefont {Schr{\o}der},\ and\ \citenamefont
  {Dyre}}]{PedersenNatCom2016}%
  \BibitemOpen
  \bibfield  {author} {\bibinfo {author} {\bibfnamefont {U.~R.}\ \bibnamefont
  {Pedersen}}, \bibinfo {author} {\bibfnamefont {L.}~\bibnamefont
  {Costigliola}}, \bibinfo {author} {\bibfnamefont {N.~P.}\ \bibnamefont
  {Bailey}}, \bibinfo {author} {\bibfnamefont {T.~B.}\ \bibnamefont
  {Schr{\o}der}}, \ and\ \bibinfo {author} {\bibfnamefont {J.~C.}\ \bibnamefont
  {Dyre}},\ }\href {\doibase 10.1038/ncomms12386} {\bibfield  {journal}
  {\bibinfo  {journal} {Nature Commun.}\ }\textbf {\bibinfo {volume} {7}},\
  \bibinfo {pages} {12386} (\bibinfo {year} {2016})}\BibitemShut {NoStop}%
\bibitem [{\citenamefont {Heyes}\ and\ \citenamefont
  {Bra{\'{n}}ka}(2015)}]{HeyesJCP2015}%
  \BibitemOpen
  \bibfield  {author} {\bibinfo {author} {\bibfnamefont {D.~M.}\ \bibnamefont
  {Heyes}}\ and\ \bibinfo {author} {\bibfnamefont {A.~C.}\ \bibnamefont
  {Bra{\'{n}}ka}},\ }\href {\doibase 10.1063/1.4937487} {\bibfield  {journal}
  {\bibinfo  {journal} {J. Chem. Phys.}\ }\textbf {\bibinfo {volume} {143}},\
  \bibinfo {pages} {234504} (\bibinfo {year} {2015})}\BibitemShut {NoStop}%
\bibitem [{\citenamefont {Heyes}, \citenamefont {Dini},\ and\ \citenamefont
  {Bra{\'{n}}ka}(2015)}]{HeyesPSS2015}%
  \BibitemOpen
  \bibfield  {author} {\bibinfo {author} {\bibfnamefont {D.~M.}\ \bibnamefont
  {Heyes}}, \bibinfo {author} {\bibfnamefont {D.}~\bibnamefont {Dini}}, \ and\
  \bibinfo {author} {\bibfnamefont {A.~C.}\ \bibnamefont {Bra{\'{n}}ka}},\
  }\href {\doibase 10.1002/pssb.201451695} {\bibfield  {journal} {\bibinfo
  {journal} {Phys. Stat. Solidi (b)}\ }\textbf {\bibinfo {volume} {252}},\
  \bibinfo {pages} {1514} (\bibinfo {year} {2015})}\BibitemShut {NoStop}%
\bibitem [{\citenamefont {Tan}, \citenamefont {Schultz},\ and\ \citenamefont
  {Kofke}(2011)}]{TanMolPhys2011}%
  \BibitemOpen
  \bibfield  {author} {\bibinfo {author} {\bibfnamefont {T.~B.}\ \bibnamefont
  {Tan}}, \bibinfo {author} {\bibfnamefont {A.~J.}\ \bibnamefont {Schultz}}, \
  and\ \bibinfo {author} {\bibfnamefont {D.~A.}\ \bibnamefont {Kofke}},\ }\href
  {\doibase 10.1080/00268976.2010.520041} {\bibfield  {journal} {\bibinfo
  {journal} {Mol. Phys.}\ }\textbf {\bibinfo {volume} {109}},\ \bibinfo {pages}
  {123} (\bibinfo {year} {2011})}\BibitemShut {NoStop}%
\bibitem [{\citenamefont {Sousa}, \citenamefont {Ferreira},\ and\ \citenamefont
  {Barroso}(2012)}]{SousaJCP2012}%
  \BibitemOpen
  \bibfield  {author} {\bibinfo {author} {\bibfnamefont {J.~M.~G.}\
  \bibnamefont {Sousa}}, \bibinfo {author} {\bibfnamefont {A.~L.}\ \bibnamefont
  {Ferreira}}, \ and\ \bibinfo {author} {\bibfnamefont {M.~A.}\ \bibnamefont
  {Barroso}},\ }\href {\doibase 10.1063/1.4707746} {\bibfield  {journal}
  {\bibinfo  {journal} {J. Chem. Phys.}\ }\textbf {\bibinfo {volume} {136}},\
  \bibinfo {pages} {174502} (\bibinfo {year} {2012})}\BibitemShut {NoStop}%
\bibitem [{\citenamefont {Khrapak}\ and\ \citenamefont
  {Morfill}(2011)}]{KhrapakJCP2011_2}%
  \BibitemOpen
  \bibfield  {author} {\bibinfo {author} {\bibfnamefont {S.~A.}\ \bibnamefont
  {Khrapak}}\ and\ \bibinfo {author} {\bibfnamefont {G.~E.}\ \bibnamefont
  {Morfill}},\ }\href {\doibase 10.1063/1.3561698} {\bibfield  {journal}
  {\bibinfo  {journal} {J. Chem. Phys.}\ }\textbf {\bibinfo {volume} {134}},\
  \bibinfo {pages} {094108} (\bibinfo {year} {2011})}\BibitemShut {NoStop}%
\bibitem [{\citenamefont {Khrapak}\ and\ \citenamefont
  {Ning}(2016)}]{KhrapakAIPAdv2016}%
  \BibitemOpen
  \bibfield  {author} {\bibinfo {author} {\bibfnamefont {S.~A.}\ \bibnamefont
  {Khrapak}}\ and\ \bibinfo {author} {\bibfnamefont {N.}~\bibnamefont {Ning}},\
  }\href {\doibase 10.1063/1.4952587} {\bibfield  {journal} {\bibinfo
  {journal} {{AIP} Adv.}\ }\textbf {\bibinfo {volume} {6}},\ \bibinfo {pages}
  {055215} (\bibinfo {year} {2016})}\BibitemShut {NoStop}%
\bibitem [{\citenamefont {Costigliola}\ \emph {et~al.}(2019)\citenamefont
  {Costigliola}, \citenamefont {Heyes}, \citenamefont {Schr{\o}der},\ and\
  \citenamefont {Dyre}}]{CostigliolaJCP2019}%
  \BibitemOpen
  \bibfield  {author} {\bibinfo {author} {\bibfnamefont {L.}~\bibnamefont
  {Costigliola}}, \bibinfo {author} {\bibfnamefont {D.~M.}\ \bibnamefont
  {Heyes}}, \bibinfo {author} {\bibfnamefont {T.~B.}\ \bibnamefont
  {Schr{\o}der}}, \ and\ \bibinfo {author} {\bibfnamefont {J.~C.}\ \bibnamefont
  {Dyre}},\ }\href {\doibase 10.1063/1.5080662} {\bibfield  {journal} {\bibinfo
   {journal} {J. Chem. Phys.}\ }\textbf {\bibinfo {volume} {150}},\ \bibinfo
  {pages} {021101} (\bibinfo {year} {2019})}\BibitemShut {NoStop}%
\bibitem [{\citenamefont {White}(1999)}]{WhiteJCP1999}%
  \BibitemOpen
  \bibfield  {author} {\bibinfo {author} {\bibfnamefont {J.~A.}\ \bibnamefont
  {White}},\ }\href {\doibase 10.1063/1.479848} {\bibfield  {journal} {\bibinfo
   {journal} {J. Chem. Phys.}\ }\textbf {\bibinfo {volume} {111}},\ \bibinfo
  {pages} {9352} (\bibinfo {year} {1999})}\BibitemShut {NoStop}%
\bibitem [{\citenamefont {Squire}, \citenamefont {Holt},\ and\ \citenamefont
  {Hoover}(1969)}]{Squire1969}%
  \BibitemOpen
  \bibfield  {author} {\bibinfo {author} {\bibfnamefont {D.}~\bibnamefont
  {Squire}}, \bibinfo {author} {\bibfnamefont {A.}~\bibnamefont {Holt}}, \ and\
  \bibinfo {author} {\bibfnamefont {W.}~\bibnamefont {Hoover}},\ }\href
  {\doibase 10.1016/0031-8914(69)90031-7} {\bibfield  {journal} {\bibinfo
  {journal} {Physica}\ }\textbf {\bibinfo {volume} {42}},\ \bibinfo {pages}
  {388} (\bibinfo {year} {1969})}\BibitemShut {NoStop}%
\bibitem [{\citenamefont {Preston}\ and\ \citenamefont
  {Wallace}(1992)}]{Preston1992}%
  \BibitemOpen
  \bibfield  {author} {\bibinfo {author} {\bibfnamefont {D.~L.}\ \bibnamefont
  {Preston}}\ and\ \bibinfo {author} {\bibfnamefont {D.~C.}\ \bibnamefont
  {Wallace}},\ }\href {\doibase 10.1016/0038-1098(92)90514-a} {\bibfield
  {journal} {\bibinfo  {journal} {Solid State Commun.}\ }\textbf {\bibinfo
  {volume} {81}},\ \bibinfo {pages} {277} (\bibinfo {year} {1992})}\BibitemShut
  {NoStop}%
\bibitem [{\citenamefont {Burakovsky}, \citenamefont {Greeff},\ and\
  \citenamefont {Preston}(2003)}]{BurakovskyPRB2003}%
  \BibitemOpen
  \bibfield  {author} {\bibinfo {author} {\bibfnamefont {L.}~\bibnamefont
  {Burakovsky}}, \bibinfo {author} {\bibfnamefont {C.~W.}\ \bibnamefont
  {Greeff}}, \ and\ \bibinfo {author} {\bibfnamefont {D.~L.}\ \bibnamefont
  {Preston}},\ }\href {\doibase 10.1103/physrevb.67.094107} {\bibfield
  {journal} {\bibinfo  {journal} {Phys. Rev. B}\ }\textbf {\bibinfo {volume}
  {67}},\ \bibinfo {pages} {094107} (\bibinfo {year} {2003})}\BibitemShut
  {NoStop}%
\bibitem [{\citenamefont {Ogata}\ and\ \citenamefont
  {Ichimaru}(1990)}]{OgataPRA1990}%
  \BibitemOpen
  \bibfield  {author} {\bibinfo {author} {\bibfnamefont {S.}~\bibnamefont
  {Ogata}}\ and\ \bibinfo {author} {\bibfnamefont {S.}~\bibnamefont
  {Ichimaru}},\ }\href {\doibase 10.1103/physreva.42.4867} {\bibfield
  {journal} {\bibinfo  {journal} {Phys. Rev. A}\ }\textbf {\bibinfo {volume}
  {42}},\ \bibinfo {pages} {4867} (\bibinfo {year} {1990})}\BibitemShut
  {NoStop}%
\bibitem [{\citenamefont {Khrapak}, \citenamefont {Chaudhuri},\ and\
  \citenamefont {Morfill}(2011)}]{KhrapakJCP2011_3}%
  \BibitemOpen
  \bibfield  {author} {\bibinfo {author} {\bibfnamefont {S.~A.}\ \bibnamefont
  {Khrapak}}, \bibinfo {author} {\bibfnamefont {M.}~\bibnamefont {Chaudhuri}},
  \ and\ \bibinfo {author} {\bibfnamefont {G.~E.}\ \bibnamefont {Morfill}},\
  }\href {\doibase 10.1063/1.3605659} {\bibfield  {journal} {\bibinfo
  {journal} {J. Chem. Phys.}\ }\textbf {\bibinfo {volume} {134}},\ \bibinfo
  {pages} {241101} (\bibinfo {year} {2011})}\BibitemShut {NoStop}%
\bibitem [{\citenamefont {Rosenberg}(2015)}]{RosenbergJPP2015}%
  \BibitemOpen
  \bibfield  {author} {\bibinfo {author} {\bibfnamefont {M.}~\bibnamefont
  {Rosenberg}},\ }\href {\doibase 10.1017/s0022377815000422} {\bibfield
  {journal} {\bibinfo  {journal} {J. Plasma Phys.}\ }\textbf {\bibinfo {volume}
  {81}},\ \bibinfo {pages} {905810407} (\bibinfo {year} {2015})}\BibitemShut
  {NoStop}%
\bibitem [{\citenamefont {Schwabe}\ \emph {et~al.}(ress)\citenamefont
  {Schwabe}, \citenamefont {Khrapak}, \citenamefont {Zhdanov}, \citenamefont
  {Pustylnik}, \citenamefont {Räth}, \citenamefont {Fink}, \citenamefont
  {Kretschmer}, \citenamefont {Lipaev}, \citenamefont {Molotkov}, \citenamefont
  {Schmitz}, \citenamefont {Thoma}, \citenamefont {Usachev}, \citenamefont
  {Zobnin}, \citenamefont {Padalka}, \citenamefont {Fortov}, \citenamefont
  {Petrov},\ and\ \citenamefont {Thomas}}]{Mierk}%
  \BibitemOpen
  \bibfield  {author} {\bibinfo {author} {\bibfnamefont {M.}~\bibnamefont
  {Schwabe}}, \bibinfo {author} {\bibfnamefont {S.}~\bibnamefont {Khrapak}},
  \bibinfo {author} {\bibfnamefont {S.}~\bibnamefont {Zhdanov}}, \bibinfo
  {author} {\bibfnamefont {M.}~\bibnamefont {Pustylnik}}, \bibinfo {author}
  {\bibfnamefont {C.}~\bibnamefont {Räth}}, \bibinfo {author} {\bibfnamefont
  {M.}~\bibnamefont {Fink}}, \bibinfo {author} {\bibfnamefont {M.}~\bibnamefont
  {Kretschmer}}, \bibinfo {author} {\bibfnamefont {A.}~\bibnamefont {Lipaev}},
  \bibinfo {author} {\bibfnamefont {V.}~\bibnamefont {Molotkov}}, \bibinfo
  {author} {\bibfnamefont {A.}~\bibnamefont {Schmitz}}, \bibinfo {author}
  {\bibfnamefont {M.}~\bibnamefont {Thoma}}, \bibinfo {author} {\bibfnamefont
  {A.}~\bibnamefont {Usachev}}, \bibinfo {author} {\bibfnamefont
  {A.}~\bibnamefont {Zobnin}}, \bibinfo {author} {\bibfnamefont
  {G.}~\bibnamefont {Padalka}}, \bibinfo {author} {\bibfnamefont
  {V.}~\bibnamefont {Fortov}}, \bibinfo {author} {\bibfnamefont
  {O.}~\bibnamefont {Petrov}}, \ and\ \bibinfo {author} {\bibfnamefont
  {H.}~\bibnamefont {Thomas}},\ }\href {\doibase 10.1088/1367-2630/aba91b}
  {\bibfield  {journal} {\bibinfo  {journal} {New J. Phys.}\ } (\bibinfo {year}
  {2020, in press}),\ 10.1088/1367-2630/aba91b}\BibitemShut {NoStop}%
\end{thebibliography}%

\end{document}